\documentclass[preprint,aps]{revtex4}
\usepackage{bm}
\usepackage{latexsym}
\usepackage{epsfig}
\input{epsf}
\def\n{{\bf n}}
\def\B{{\bf B}}
\def\L{{\bf L}}
\def\r{{\bf r}}
\def\s{{\bf s}}

\def\h{{\bf h}}
\def\om{\vec{\omega}}
\begin{document}

\title{ 
Can the rotation of the dark matter halo of our galaxy be detected
through its effect on the cosmic microwave background polarisation?}
\author{Partha Nag{\footnote{Present address : Tata Institute of
Fundamental Research, Homi Bhabha Road, Mumbai 400005, India}}, 
Somnath Bharadwaj and Sayan Kar}
\email{sayan@cts.iitkgp.ernet.in, somnath@cts.iitkgp.ernet.in}
\affiliation{Department of Physics and Meteorlogy, \&  Centre for
  Theoretical Studies  
\\ 
Indian Institute of Technology, Kharagpur 721 302, WB, India}
\begin{abstract}
The proposition that dark matter halos possess angular
momentum, though widely accepted, is a theoretical prediction which
has, till date, not been observationally verified.  The gravi-magnetic
field produced by a rotating gravitating object is possibly the only
direct consequence  of the angular momentum of dark matter halos. The
gravitational Faraday rotation produced by the dark matter halo of our
galaxy is present in all astronomical observations. A detection of 
the imprint of this effect on the all-sky cosmic microwave background 
radiation polarisation pattern would directly probe the angular
momentum of the dark matter halo of our Galaxy. We have calculated the
expected gravitational Faraday rotation which turns out to be a 
$\left (\frac{v_c}{c}\right )^3$ effect, $v_c$ being the rotational
speed of our halo. The predicted gravitational Faraday effect rotation
angles are less than $1^{''}$, implying that this effect, though present,
is too small to be detected. 
\end{abstract}


\maketitle

\section{Introduction}
Galaxies and clusters of galaxies are believed to be embedded in dark
matter halos whose masses are significantly larger than those of the
visible baryonic components, namely, stars and  gas.
Determining  the nature of the dark matter is currently one of the
most challenging  problems in astrophysics  
{\cite{dark}}. The dark matter halo makes its presence felt only
through its gravitational field which, therefore,
merits particular attention. Several proposals on 
modeling the dark matter halo can be found in {\cite{halogravf,partdark}.    

The dark matter halos are predicted to have angular momentum. In the
hierarchical structure formation scenario the  halos acquire angular
mementum through tidal interactions of the halos with their
surroundings {\cite{peebles}}.  The angular momentum is usually
quantified using the dimensionless spin parameter $\lambda=J \mid E
\mid^{1/2}/G M^{5/2}$ where $E$ is the total energy of the halo and
$M$ is its mass. The spin parameter $\lambda$ of a halo is essentially
the ratio of its angular momentum to that needed for rotational 
support. Simulations show $\lambda$ to be in the range $0.02- 0.1$ 
{\cite{vitvitska}}. 

In Newtonian gravity the gravitational field of an object does not
depend on whether the object is rotating or not, and the angular
momentum of a dark matter halo does not manifest itself in its
gravitational field. Interestingly, this is not true in Einstein's
theory of gravity where the gravitational field of a rotating object
is different from that of a static one. The angular momentum
manifests itself itself through the gravomagnetic effect \cite{gravomag},
which is analogous to the magnetic field produced by a rotating
charge. 

The gravomagnetic effect is possibly the only signature of the angular
momentum of a dark matter halo, and any detectable  consequence of
this effect holds the only hope of  measuring the angular momentum of
dark matter halos. In this paper, for a rotating dark matter halo, we  
calculate the gravomagnetic field both inside and outside the
halo. The gravitational field inside the halos of galaxies and
clusters of galaxies is weak, and we  adopt the weak
field limit of Einstein's theory in our calculation. 

The Faraday effect is a well known manifestation of magnetic fields on
astronomical scales, and the gravo-magnetic field 
produced by rotating
dark matter halos will produce an analogous gravitational Faraday
effect \cite{gravofarad}. This will cause the plane of polarization of 
linearly polarized electro-magnetic radiation to rotate as it propagates
through the gravitational field of a rotating dark-matter halo. 
The gravitational Faraday rotation of the dark matter halo of our
Galaxy will be present in all astronomical observations carried out
from our location inside the halo. In particular, the gravitational
Faraday effect will effect the observed polarisation pattern of the
Cosmic Microwave Background Radiation (CMBR). In this paper, we
calculate the expected signature of the gravitational Faraday effect of
the dark matter halo of our Galaxy and investigate if this will be
detectable in the large angular scale CMBR polarisation pattern
\cite{cmbrpol}.   
 
A brief outline of the paper follows. In Section 2 we calculate the
metric for a rotating dark matter halo and use this to determine the
gravo-magnetic field both inside and outside the halo. In Section
3 we calculate the Faraday effect that will be produced by the
gravo-magnetic field of the halo and estimate its effect on the CMBR
polarisation pattern as seen by an observer inside the halo. In
Section 4 we present results and conclusions. 

It should be noted that throughout this paper we have restricted our
analysis to the dark matter halo of a spiral galaxy. A similar
analysis can also be applied to clusters of galaxies {\cite{cluster}}
without major modifications, but this has not been considered here. 

\section{The Gravomagnetic Field}
The gravitational field associated with galaxy  dark matter halos is
weak and it is justified to describe it using  a metric 
\begin{equation}
g_{\mu \nu}=\eta_{\mu \nu} + h_{\mu \nu}
\label{eq:1}
\end{equation}
which deviates only slightly from the flat space-time Minkowski metric
$\eta_{\mu \nu}$ {\it ie.} $\mid h^{\mu}_{\nu} \mid \ll 1$.
It may be noted that we adopt the  signature $(-,+,+,+)$ for the
metric and 
the weak field (linearised) Einstein equations in the
Lorentz gauge  {\cite{shutz}} are given by 
\begin{equation}
\Box H^{\alpha \beta}=-\frac{16 \pi G}{c^4} T^{\alpha \beta}
\label{eq:2}
\end{equation}
where 
\begin{equation}
H^{\alpha \beta}=h^{\alpha \beta}-\frac{h}{2}\eta^{\alpha \beta}.
\label{eq:3}
\end{equation}

To begin with, we ignore the rotation and consider a
static, spherically symmetric halo. The most general static, spherically
symmetric  spacetime  has metric coefficients of the form  
$h_{00}=-2 \Phi(r)$, $h_{i0}=0$ and $h_{ij}=-2 [\Phi(r)+\Psi(r)]
\delta_{ij}$ where $\Phi(r)$ and $\Psi(r)$ are two arbitrary functions
of the radial distance $r$. Further, if we assume that  the dark
matter does not have relativistic pressure ({\it ie.} $T_{00}=\rho c^2$ is
the only non-zero component of the energy momentum tensor) it then 
follows from eq. (\ref{eq:2}) that $\Psi(r)=0$.  The metric now 
contains only  one unknown function $\Phi(r)$ which is
the usual Newtonian gravitational potential divided by  $c^{2}$.
For spiral galaxies $\Phi(r)$ can be determined from HI rotation
curves. Assuming the HI to be rotating in circular orbits, the
geodesic equation reduces to 
\begin{equation}
\frac{d \Phi}{d \, r}=\frac{1}{r}\frac{v_c^2(r)}{c^2}
\label{eq:4}
\end{equation}
where $v_c(r)$ is the velocity of the HI cloud at a distance $r$
fromt the center of the galaxy. The outer parts of 
spiral galaxies usually show a flat rotation curve {\it ie.} $v_c$ is
a constant {\cite{rc}}. For this, the potential, up to an arbitrary constant of 
integration,  is 
\begin{equation}
\Phi(r)=\frac{v^2}{c^2} \ln r  \,.
\label{eq:5}
\end{equation}

Using this in the Einstein's equation (\ref{eq:2}) gives the total density
in the outer parts of spiral galaxies to be $\rho(r)=v^2/4 \pi G
r^2$. This density is usually significantly larger than the density of
the visible matter and this is the dark matter problem. The density
profile in the inner parts of galaxies is not very well determined and
it is still an issue of debate. For the purposes of this paper we
assume the density profile 
\begin{equation}
\rho(r)=\frac{v^2}{4\pi G(r^2+r_0^2)} 
\label{eq:6}
\end{equation}
for the dark matter halo throughout the galaxy. Here $r_0$ is the core
radius and it is assumed that the Dark Matter halo is of size $R$ {\it
  ie.} $\rho(r)=0$ for $r>R$. 

We now incorporate the rotation of the  halo. Retaining the  spherical
density profile given by eq. (\ref{eq:6}) 
we assume that the halo is  made up of
spherical shells, each shell executing rigid rotation with
angular velocity $\om(r)$ whose magnitude and direction can vary
  from shell to shell.  It is also assumed that the rotational
  velocities $\vec{v}=\om \times \r$ are non-relativistic ($\mid
  \vec{v}/c \mid \ll 1$). The change introduced by the rotation is
  that we now have a non-zero time-space component of the
  energy-momentum tensor   $T^{0 i}=\rho c v^{i}$. This results in 
  non-zero values for the metric coefficients $\h=h_{oi}$ whose values
  have to  be determined through 
\begin{equation}
\nabla^2 \h(\r)=\frac{16\pi G }{c^3}\rho(r)({\om} \times \r) \,.
\label{eq:7}
\end{equation}

The solution for  ${\bf h}$ is given by
\begin{equation}
{\bf h}({\bf s})=-\frac{4G}{c^3} \int\frac{\rho(r)({\bf \om(\r)} \times {\bf
    r})}{|{\bf 
    s}-{\bf r}|} d\tau 
\label{eq:8}
\end{equation}

The integral in eq. (\ref{eq:8}) can be simplified and written as a
sum of two parts as 

\begin{equation}
{\bf h}({\bf s})=-\frac{16\pi G}{3 c^3}\left( \frac{1}{s^2}
\int_{0}^{s} r^4 \rho(r)~\om(r)~dr+s\int_{s}^{R}
r~\rho(r)~\om(r) \right)\times\frac{\bf s}{s}
\label{eq:9}
\end{equation}

where the first integral is the contributionfrom the region interior
to the point $\s$ where $\h$ is being calculated and the second
integral is for the exterior region. 

The angular momentum of a shell of radius $r$ and thickness $dr$ is 
\begin{equation}
d \L(r) =\frac{8 \pi}{3} \om(r) \rho(r) r^4 dr
\end{equation}
Using this, we can write eq. (\ref{eq:8}) as  
\begin{equation}
{\bf h}({\bf s})=-\frac{2G}{c^3}\left( \frac{\L_s}{s^2} +s \int_{s}^{R} 
 \frac{d \L(r)}{r^3}  \right)\times\frac{\bf s}{s}
\label{eq:10}
\end{equation}

Here $\L_s$ denotes the total angular momentum of all the shells
interior to $\s$. The first term in  eq. (\ref{eq:10}) for $\h(\s)$ is
the contribution from the total angular momentum interior to $\s$ and the
term involving the integral is the contribution from the angular
momentum of the shells exterior to $\s$.
     
It is worth  noting that only the first term contributes outside the halo 
and relates $\h(\s)$ to the angular momentum of the whole halo
\begin{equation}
{\bf h}({\bf s})=2\frac{{\bf L}\times{\bf s}}{s^3}
\end{equation}

Calculating the geodesic equation for a non-relativistic test particle
 moving in the gravitation field of the halo we have 
\begin{equation}
\frac{d {\bf v}}{d t}=-{\bf \nabla} (c^2 \Phi)+{\bf v} \times ({\bf
    \nabla} \times \h \, c)
\label{eq:12}
\end{equation}
where we see that we can identify $c \h$ as the equivalent of the
vector potential ${\bf A}$ in electro-magnetism, and we can identify
$\B_g= 
({\bf\nabla} \times \h \, c)$ as the gravi-magnetic field. This will
affect all motion inside the dark matter halo and it is possible that
there may be a detectable effect of this force. We do not consider
this possibility here but instead we focus on the gravitational
Faraday effect which is  a different manifestation of  the
gravi-magnetic field. 

\section{Gravitational Faraday Effect}

Having obtained the gravi--magnetic field let us now see if we can
predict any observable effect.     
It is known  that the plane of polarisation of a light ray passing through 
a region with a gravimagnetic field may get rotated due to
the gravitational Faraday effect. This is similar to the magnetooptic 
Faraday effect {\cite{gravofarad}} except that the gravitational analogue is
achromatic (wavelength independent). The rotation of the plane of
polarisation is a consequence of the parallel 
transport of the polarisation vector along the path of the light ray. 
The resultant rotation of the plane of polarisation is
given by (see paper by Sereno in {\cite{gravofarad}}) the expression:  
\begin{equation}
\Omega=\frac{1}{2}\int{\bf B}_g \cdot  d{\bf l}
\end{equation} 
where the integral is along the trajectory of the light ray. For the CMBR
 being observed from inside our Galaxy we  consider a light ray
 propagating along the line of sight starting from infinity (actually
 the last scattering surface)  and ending at the observer. 

The quantity of interest from the observational point of view is the
difference in the rotation angle for CMBR  along different lines of
sight. There will be no detectable effect if the rotation angle were
the same along all lines of sight.  We consider CMBR photons arriving
along two different lines of sight say $\n_1$ and $\n_2$. The light
rays reaching us along the two different lines of sight will traverse
two different paths and 
the rotation angles for their planes of polarisation will be 
given by
\begin{equation}
\Omega_{1}=\frac{1}{2}\int {\bf B_g} \cdot d{\bf l}_{1}
\end{equation}
and
\begin{equation}
\Omega_{2}=\frac{1}{2}\int {\bf B}_g \cdot d{\bf l}_{2}
\end{equation}
where the integrals are still from infinity to the observer but along the 
two different paths. Now consider a path joining the two points at
infinity. This third path lies entirely at infinity where the gravimagnetic 
field vanishes. Thus the integral over this third path vanishes. 
Thus if we consider the difference of the two rotation angles $\Omega_{1}$
and $\Omega_{2}$ we can safely add to it the integral along the third path at 
infinity since this does affect the net result. Thus the 
difference of the two angles yields a closed integral of the form    
\begin{equation}
\Omega_{1}-\Omega_{2}=\frac{1}{2}\oint{\bf B}_g \cdot d{\bf l}
\end{equation}
Applying Stokes' theorem we obtain 
\begin{equation}
\Omega_{1}-\Omega_{2}=\frac{1}{2}\int {\bf \nabla} \times{\bf B}_g
  \cdot d{\bf 
  s}=\frac{c}{2}\int {\bf \nabla}\times( {\bf \nabla}\times{\bf h})
  \cdot d{\bf s}  
\end{equation}
where the integral is over the surface enclosed by the two referred paths.
Note that we are working in the Lorentz gauge where the divergence of the 
vector ${\bf h}$ vanishes. Also the Laplacian of this vector is 
given as

\begin{equation}
\nabla^{2}{\bf h}=-\frac{16 \pi G}{c^3} \rho(r)\left ({\bf
  \omega}\times{\bf r} \right ) 
\end{equation}
Thus the expression for the difference in the two rotation angles becomes 
\begin{equation}
\Omega_{1}-\Omega_{2}=\frac{8\pi G}{c^2} \int\rho(r)({\bf
  \omega}\times{\bf r}) \cdot d{\bf s}=\frac{8\pi G}{c^2} \int{\bf
  j} \cdot d{\bf s}
\end{equation}
where the quantity ${\bf j}=\rho {\bf v}$  is the matter current density.
 
Thus we see that the angle difference is proportional to nothing but the 
matter flux across the plane enclosed by the two referred
paths. Only a finite portion of this surface lying within the
spherical halo  contributes to this.

\begin{figure}[hbt]
\includegraphics[angle=0,width=0.2\textwidth]{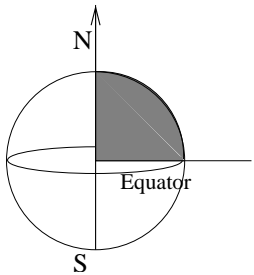}
\caption{This shows the rotating halo. The shaded region shows the
  surface trhough which the matter flux is to be calculated.}
\label{fig:1}
\end{figure}

For simplicity we assume that all the shells in the halo have a common
rotation axis. Relative to the center of the halo, we shall refer to
one direction along the rotation axis as the North pole and the
opposite direction as the South pole, and the plane perpendicular to
the rotation axis as the equator (Figure \ref{fig:1}) . From the
symmetry of the problem 
we expect $\B_g$ to be perpendicular to the light ray for a lines of
sight along the equator. Hence  we do not expect any gravitational
Faraday effect along the equator.  Further the effect is expected to
be maximum at the poles, the direction of the rotation being opposite
in the two hemispheres and the magnitude of the rotation increasing
with elevation away from the equatorial plane. 

We next estimate the rotation of the polarisation angle along one of
   the poles by calculating the matter flux through a surface shown in
   Figure \ref{fig:1}. 
Using the density profile in eq. (\ref{eq:6}) and assuming that the
all shells of the halo rotate with the same velocity as the HI in the
spiral galaxy embedded in the halo we have 

\begin{equation}
\mid \Omega_{N} \mid =\pi  \frac{v_c^3}{c^3} \int^R_{r_0} \frac{r \,
  dr}{r^2 +r_0^2}
\label{eq:20}
\end{equation}
where $\Omega_N$ is the rotation angle along the North
pole. Calculating this we have 

\begin{equation}
\mid \Omega_{N} \mid =\pi  \frac{v_c^3}{c^3} \ln\left( \frac{R}{r_0}
\right) 
\label{eq:21}
\end{equation}
where we have assumed $r_0 \ll R$. 

\section{Result and Conclusion}

Galaxy  rotational velocities are trypically $\sim 200 {\rm km/s}$ for
which $({v_c}/c)^3\sim 3 \times 10^{-10}$. The only hope for a substantial
rotation is through the factor $\ln(R/r_0)$. The size of the core
radius and the extent of the halo are both largely not well
determined. Assuming that the core radius is of the order of a few
${\rm pc}$ and that the halo extends to a few ${\rm Mpc}$ we see that
the factor $\ln(R/r_0)$ will have a value in the range $10-15$.  
It then follows that the rotation will be less than $1^{''}$ which is
too small to  have a detectable signature in the CMBR. 

It should be noted that the gravitational Faraday effect is of the
order of $({v_c}/c)^3$ and hence it is small for most non-relativistic
situations.

 The analysis carried out here depends crucially on the
fact that the density profile falls as $r^{-2}$ at large $r$, and hence
the matter flux (and  the rotation angle) increases only
logarithmically with the size of the halo. It is possible that our
conclusions would be different (and the rotation detectable) if the
density profile were to be shallower.  

A worthwhile future investigation will be to study gravitational lensing 
in the line element we have obtained for the rotating galaxy halo. 
The existent literature {\cite{sereno}} 
seems to suggest that the deflection angle
will be significantly different because of gravomagnetic effects.
It remains to be seen what lies in store for us if the deflection
is worked out for the metric we have derived here.

\newpage

\end{document}